\documentclass[aps,pra,floatfix,twocolumn,showpacs,10pt]{revtex4-1}
\usepackage{amssymb, amsmath,color,mciteplus,graphicx,subfigure}
\usepackage{calc,epsfig,epstopdf,color,mciteplus,bm,mathrsfs}
\usepackage{times}
\usepackage{float}
\usepackage{lipsum}

\begin{document}

\title{Scalable nonadiabatic holonomic quantum computation on a superconducting qubit lattice}

\author{Li-Na Ji}
\author{Tao Chen}
\author{Zheng-Yuan Xue} \email{zyxue83@163.com}
\affiliation{Guangdong Provincial Key Laboratory of Quantum Engineering and Quantum Materials, GPETR Center for Quantum Precision Measurement, and School of Physics and Telecommunication Engineering,  South China Normal University, Guangzhou 510006, China}

\date{\today}

\begin{abstract}
  Geometric phase is an indispensable element for achieving robust and high-fidelity quantum gates due to its built-in noise-resilience feature. However, due to the complexity of manipulation and the intrinsic leakage of the encoded quantum information to non-logical-qubit basis, the experimental realization of universal nonadiabatic holonomic quantum computation is very difficult. Here, we propose to implement scalable nonadiabatic holonomic quantum computation with decoherence-free subspace encoding on a two-dimensional square superconducting transmon-qubit lattice, where only the two-body interaction of neighboring qubits, from the simplest capacitive coupling, is needed. Meanwhile, we introduce qubit-frequency driving to achieve tunable resonant coupling for the neighboring transmon qubits, and thus avoiding the leakage problem. In addition, our presented numerical simulation shows that high-fidelity quantum gates can be obtained, verifying the advantages of the robustness and scalability of our scheme. Therefore, our scheme provides a promising way towards the physical implementation of robust and scalable quantum computation.
\end{abstract}

\maketitle

\section{Introduction}
Quantum computation, due to the characteristic of coherent superposition of quantum states, can speed up the processing of certain complex problems, such as factoring large integers and searching unsorted databases. However, a quantum system is inevitably coupled to its surrounding environment, resulting in irreversible destruction of the encoded quantum information, leading to errors in manipulating a quantum system. Meanwhile, errors during the control over a quantum system will also introduce additional imperfection to a quantum gate operation. Thus, how to achieve high-fidelity quantum gates on quantum systems becomes the key problem in the realization of scalable quantum computation.

Geometric phases \cite{GP1,GP2}, determined by the global properties of the evolution paths, becoming the indispensable elements for achieving robust and high-fidelity quantum gates, due to the built-in noise-resilience feature \cite{robust1,robust2,robust3,robust4}. It is well known that holonomic quantum computation \cite{HQC1} based on non-Abelian geometric phase can be achieved by adiabatic cyclic evolution \cite{Duan,HQC2,HQC3,HQC4,HQC5}. Unfavourably, the adiabatic proposals require the target quantum systems to be exposed to their environment for a long time, thereby decoherence effect will lead to considerable influence, which obliterates the advantage of the geometric phase. To speed up the quantum gate operation, nonadiabatic holonomic quantum computation (NHQC) \cite{NHQC1,DFSNHQC2} has been proposed to construct universal quantum gates. Then, various NHQC schemes have been proposed theoretically \cite{NHQC2,NHQC3,NHQC4,NHQC5,NHQC6,NHQC7,NHQC8,NHQC9,NHQC10,NHQC11,NHQC12,NHQC13} and demonstrated experimentally in many systems \cite{exp7,exp8,exp9,exp10,exp11,exp12, exp13,exp14,exp15,exp16,exp17,exp18}. However, because of the complexity of experimental manipulation and the intrinsic leakage of the encoded quantum information out of the logical-qubit basis, the experiment of high-fidelity universal NHQC, in particularly the nontrivial two-qubit gates, is very difficult.

Meanwhile, to suppress collective dephasing noise which is the main source of decoherence, schemes with decoherence-free subspace (DFS) encoding \cite{DFS1,DFS2,DFS3} have been proposed.
Thus, NHQC based on DFS encoding \cite{DFSNHQC2,DFSNHQC3,DFSNHQC4,DFSNHQC5,DFSNHQC6,DFSNHQC7,DFSNHQC8} can combine the operational robust feather of geometric phase and decoherence resilience of DFS encoding. However, due to the need for precise interactions among the multiple quantum systems, the experiment of NHQC in DFS faces great challenges.

Here, we propose a practical scheme to implement universal NHQC in a DFS on a scalable two-dimensional (2D) square lattice with capacitive coupled superconducting qubits, which removes the above-mentioned difficulties. The key merit of our scheme is that it only involves the two-body interaction of the neighboring transmons. Meanwhile, we introduce additional qubit-frequency driving to implement tunable coupling for the neighboring transmons in an all-resonant way, thus avoiding the leakage problem and can result in the robust and high-fidelity universal quantum holonomic gates in a simple setup. Therefore, our scheme provides a promising method to achieve high-fidelity geometric manipulation for robust and scalable solid-state quantum computation.

\section{Tunable Interaction}

\begin{figure}[tbp]
  \centering
  \includegraphics[width=0.9\linewidth]{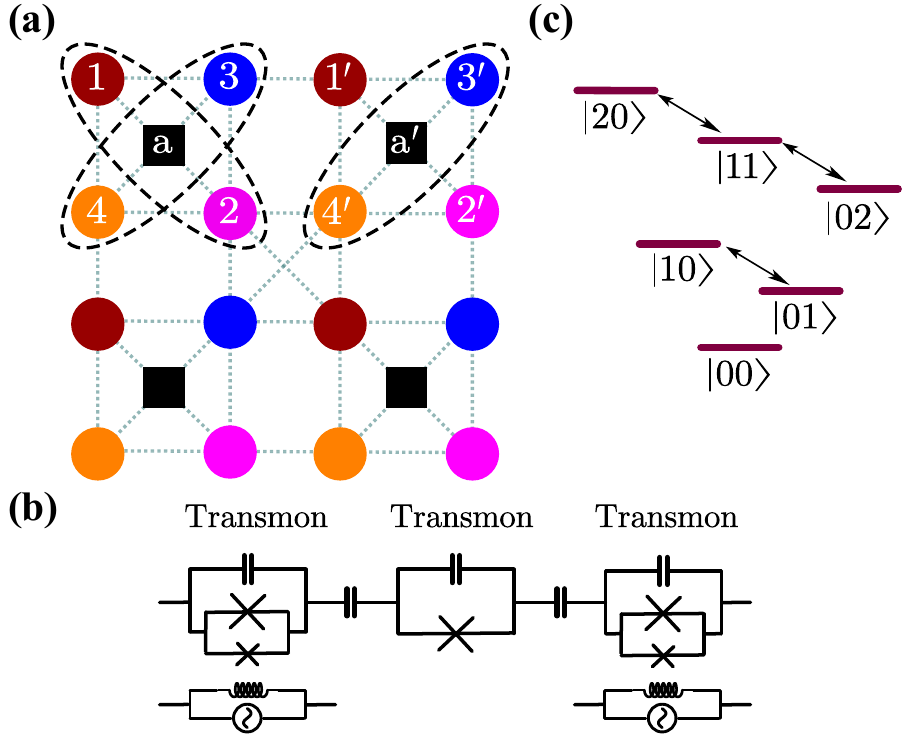}
  \caption{Our proposed setup. (a) A scalable 2D square lattice, where each ellipse denotes a DFS encoded logical qubit, which consists of a pair of transmons with different frequencies connected by an auxiliary transmon, the corresponding circuits are shown in (b). (c) Energy spectrum for two parametrically tunable coupled transmons, where single- and two-excitation subspaces can be used to achieve single- and two-logical-qubit holonomic gates, respectively.}\label{Fig1}
\end{figure}

We first review the used ac magnetic flux induced parametric tunable coupling, which can be implemented between a qubit and a quantum bus \cite{flux-driven0,flux-driven1,tc1}, or two qubits \cite{flux-driven2,flux-driven3,tc2,lix}. As shown in Fig. \ref{Fig1}(a), two neighboring transmon qubits $\textrm{T}_i$ and $\textrm{T}_j$ are capacitive coupled, where subscripts $(i,j)$ indicates the position of neighboring transmons on a 2D square superconducting transmon-qubit lattice, e.g., $(i,j)=(1,a), (2,a)$, etc.  Assuming $\hbar=1$ hereafter, the Hamiltonian of the coupled system is
\begin{eqnarray}
\label{Eq1}
\mathcal{H}_c&=&\sum_{l=i,j} \sum_{k=1}^2 [k\omega_l-(k-1)\alpha_l]\chi^k_l\notag\\ &&
 +  \left[ g_{{ij}} \prod_{l=i,j} \left(\sum_{k=1}^2 \lambda_k\sigma^k_l \right) +\mathrm{H.c.}\right],
\end{eqnarray}
where $g_{{ij}}$ is the coupling strength; $\chi^{k}=|k\rangle\langle k|$ and $\sigma^k=|k-1\rangle\langle k|$ are the projector for the $k$th level and the standard lower operator, respectively; the associated transition frequency is $[k\omega-(k-1)\alpha]$ with $\alpha$ being the intrinsic anharmonicity of transmon; and $\lambda_k=\sqrt{k}$ weighs the relative strength of the $|k\rangle \leftrightarrow |k-1\rangle$ transition. To consider the leakage effect, we have taken the third energy level of the transmons into account.

However, for capacitive coupled transmon qubits, the coupling strength are fixed and difficult to adjust. Meanwhile, the frequency difference $\Delta_i=\omega_i-\omega_j$ of two qubits is fixed and nonzero in general, making it impossible to ensure that they are resonant coupled when they both work in their optimal points. Thus, to realize the discretionarily controllable manipulation between two neighboring transmons, for one of them, such as $\textrm{T}_i$, we introduce an additional qubit-frequency driving, in the form of $\omega_i(t)=\omega_i+\varepsilon_i \sin(\nu_i t+\phi_i)$, which can be experimentally achieved by biasing the transmon with an ac magnetic flux in a particular dc bias working point \cite{flux-driven3,tc2}, the circuit details are shown in Fig. \ref{Fig1}(b). Moving into the interaction picture and using the Jacobi-Anger identity of
\begin{eqnarray}
\label{EqJai}
&&\exp[-\emph{i}\beta_i \cos(\nu_i t+\phi_i)] \notag \\
&&\quad \quad \quad \quad=\sum^{\infty}_{n=-\infty} (-\text{i})^n J_n(\beta_i) \exp[-\text{i}n(\nu_i t+\phi_i)], \notag
\end{eqnarray}
the transformed Hamiltonian can be written as
\begin{eqnarray}
\label{Eq2}
\mathcal{H}_{ij}&=& \sum_{n=-\infty}^{+\infty}(-\text{i})^n J_n(\beta_i) g_{ij}\left\{ |10\rangle_{ij}\langle 01|e^{\text{i}\Delta_i t}e^{-\text{i}n(\nu_it+\phi_i)}\right. \notag \\
&+&\left.\sqrt{2}|11\rangle_{ij}\langle 02|e^{\text{i}(\Delta_i+\alpha_j) t}e^{-\text{i}n(\nu_it+\phi_i)} \right. \notag \\
&+&\left.\sqrt{2}|20\rangle_{ij}\langle 11|e^{\text{i}(\Delta_i-\alpha_i) t}e^{-\text{i}n(\nu_it+\phi_i)}\right\}+\mathrm{H.c.},
\end{eqnarray}
where $J_n(\beta_i)$ are Bessel functions of the first kind, and $\beta_i=\varepsilon_i/\nu_i$. Then neglecting the high-order oscillating terms by the rotating-wave approximation, we find that parametrically tunable resonant coupling in the single- and/or two-excitation subspaces can be both achieved by only modulating the qubit-frequency driving parameters $(\varepsilon_i, \nu_i)$, the corresponding energy level diagram is shown in Fig. \ref{Fig1}(c).

\section{Single-logical-qubit holonomic gates}
We now proceed to implement universal NHQC with DFS encoding on a scalable 2D square superconducting transmon-qubit lattice.
For the case of a single-logical qubit, with minimum resource requirement, we here only use two transmons as a logical unit to encode a logical qubit, i.e.,
\begin{eqnarray}
\label{Eq3}
S_1= \textrm{Span}\{|10\rangle_{12}=|0\rangle_L, \ \ |01\rangle_{12}=|1\rangle_L \}.
\end{eqnarray}

The corresponding control Hamiltonian can be defined as
$\mathcal{H}_S=\mathcal{H}_{1\textrm{a}}+\mathcal{H}_{2\textrm{a}}$,
where transmon $\textrm{T}_\textrm{a}$ is considered as an auxiliary element to synchronously achieve the parametrically tunable coupling interaction with its two neighboring transmons $\textrm{T}_1$ and $\textrm{T}_2$, which are driven by ac magnetic fluxes.
By modulating the qubit-frequency driving parameters $\nu_1=\Delta_1$ and $\nu_2=\Delta_2$, see Eq. (\ref{Eq2}), we can obtain the following effective resonant interaction Hamiltonian:
\begin{eqnarray}
\label{Eq5}
\mathcal{H}_1=g'_{1\textrm{a}}e^{-\text{i}\phi'_1}|10\rangle_{1\textrm{a}}\langle 01| +g'_{2\textrm{a}}e^{-\text{i}\phi'_2}|10\rangle_{2\textrm{a}}\langle 01|+ \mathrm{H.c.},\ \
\end{eqnarray}
where $g'_{l\textrm{a}}=J_1(\beta_l) g_{l\textrm{a}}$ and $\phi'_l=\phi_l+{\pi}/{2}$.  In addition, the interaction form of different subspaces is shown in Eq. (\ref{Eq2}), the manipulations of the single-logical-qubit states are limited in the single-excitation subspace of Hamiltonian $\mathcal{H}_S$, the resonant modulation of which is independent of the anharmonicity of transmon, thus the level leakage to the multi-excitation subspaces can be directly eliminated.

Then by defining $g=\sqrt{g'^2_{1\textrm{a}}+g'^2_{2\textrm{a}}}$, $\theta=2\tan^{-1}(g'_{2\textrm{a}}/ {g'_{1\textrm{a}}})$ and $\phi=\phi'_2-\phi'_1$, the Hamiltonian $\mathcal{H}_1$ in the auxiliary qubit basis $\{|0\rangle_{\textrm{a}}, |1\rangle_{\textrm{a}}\}$ can be reduced to
\begin{eqnarray}
\label{Eq5}
\mathcal{H}_{L_1}=g\left(
\begin{array}{cccc}
 0 & K e^{-\text{i}\phi'_1} \\
 K^{\dagger}e^{\text{i}\phi'_1} & 0
\end{array}
\right),
\end{eqnarray}
with
\begin{eqnarray}
\label{Eq6}
K=\left(
\begin{array}{cccc}
 0 & 0 & 0 & 0\\
 \sin\frac{\theta} {2}e^{-\text{i}\phi} & 0 & 0 & 0\\
 \cos\frac{\theta} {2} & 0 & 0 & 0\\
 0 & \cos\frac{\theta} {2} & \sin\frac{\theta} {2}e^{-\text{i}\phi} & 0
\end{array}
\right),
\end{eqnarray}
in a four-dimensional basis $\{|00\rangle_{12},$ $|1\rangle_{L}, |0\rangle_{L}, |11\rangle_{12}\}$. To analyze the evolution of logical-qubit subspace $S_1$, we decompose matrix $K$ in the form of $K=XYZ^{\dagger}$ with
\begin{eqnarray}
\label{Eq}
X&=&\left(
\begin{array}{cccc}
 1 & 0 & 0 & 0\\
 0 & \cos\frac{\theta} {2} & 0 & \sin\frac{\theta} {2}e^{-\text{i}\phi}\\
 0 & -\sin\frac{\theta} {2}e^{\text{i}\phi} & 0 & \cos\frac{\theta} {2}\\
 0 & 0 & 1 & 0
\end{array}
\right), \quad
Y=\left(
\begin{array}{cccc}
 0 & 0 & 0 & 0\\
 0 & 0 & 0 & 0\\
 0 & 0 & 1 & 0\\
 0 & 0 & 0 & 1
\end{array}
\right), \notag \\
Z&=&\left(
\begin{array}{cccc}
 0 & 0 & 0 & 1\\
 0 & \sin\frac{\theta} {2}e^{-\text{i}\phi} & \cos\frac{\theta} {2}  & 0\\
 0 & -\cos\frac{\theta} {2} & \sin\frac{\theta} {2}e^{\text{i}\phi} & 0\\
 1 & 0 & 0 & 0
\end{array}
\right).  \notag
\end{eqnarray}
Substituting the matrix decomposition into Eq. (\ref{Eq5}), the corresponding time evolution operator can be obtained as
\begin{eqnarray}
\label{Eq}
U_1(t)=
\left(
\begin{array}{cccc}
 X\cos(a_{t}Y)X^{\dagger} & -\text{i}X\sin(a_{t}Y)Z^{\dagger}e^{-\text{i}\phi'_1} \\
 -\text{i}Z\sin(a_{t}Y)X^{\dagger}e^{\text{i}\phi'_1} & Z\cos(a_{t}Y)Z^{\dagger}
\end{array}
\right),\notag
\end{eqnarray}
where $a_{t}=gt$.

\begin{figure}[tbp]
  \centering
  \includegraphics[width=0.95\linewidth]{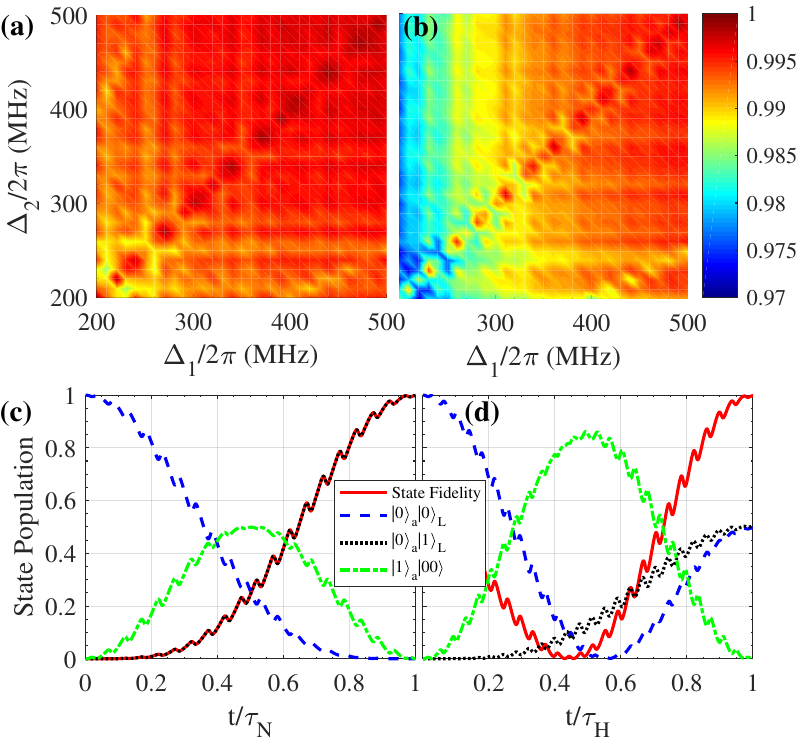}
  \caption{The gate fidelities as functions of the qubit frequency differences $\Delta_1$ and $\Delta_2$, the results of the NOT and Hadamard gates are shown in (a) and (b), respectively, with the same color bar. State populations and the state fidelity dynamics of the NOT (c) and Hadamard gate (d) operations with the initial state being $|0\rangle_\textrm{a}|0\rangle_L$.}\label{Fig2}
\end{figure}

Generally, when $a_{\tau}=\pi$ at the final time $\tau$, arbitrary single-logical-qubit holonomic gates can be achieved by two sequential evolutions. To avoid the longer gate time, we here divide a single-loop evolution with duration $\tau$ into two equal segments $[0,{\tau}/ {2}]$ and $[{\tau} /{2},\tau]$, where we only change the phase $\phi'_1$ to $\phi'_1+\pi+\gamma$ at ${\tau}/{2}$. Then, the evolution operator is
\begin{eqnarray}
\label{Eq7}
U_1(\tau)&=&U_1\left(\tau,\frac{\tau} {2}\right)U_1\left(\frac{\tau} {2},0\right) \\
&=& |0\rangle_{\textrm{a}}\langle 0|\otimes X[\cos^2(a_{\frac{\tau} {2} }Y)+\sin^2(a_{\frac{\tau} {2} }Y) e^{-\text{i}\gamma}]X^{\dagger} \notag \\
&+& |1\rangle_{\textrm{a}}\langle 1|\otimes Z[\cos^2(a_{\frac{\tau} {2} }Y)+\sin^2(a_{\frac{\tau} {2} }Y) e^{\text{i}\gamma}]Z^{\dagger}, \notag
\end{eqnarray}
where by restricting the initial state of the auxiliary transmon $\textrm{T}_\textrm{a}$, the reduced evolution operators in different subspaces can be achieved. For example, if we initially prepare the auxiliary transmon in its ground state $|0\rangle_{\textrm{a}}$, the evolution operator within the single-logical-qubit subspace $\{|0\rangle_{L}, |1\rangle_{L}\}$ will be
\begin{eqnarray}
\label{Eq8}
U_{L_1}(\tau)=\cos\frac{\gamma} {2}-\text{i}\sin\frac{\gamma} {2}\left(
\begin{array}{cccc}
 \cos\theta & \sin\theta e^{-\text{i}\phi} \\
 \sin\theta e^{\text{i}\phi} & -\cos\theta
\end{array}
\right).
\end{eqnarray}
In this way, arbitrary single-logical-qubit holonomic gates can be achieved by the selection of parameter $(\theta,\gamma,\phi)$ in a single-loop scenario. For example, by setting $\theta={\pi}/{2}$ and ${\pi}/{4}$ with the same $\gamma=\pi$ and $\phi=0$, the NOT and Hadamard gates can be obtained, respectively.

Note that the above processes can be identified as nonadiabatic holonomy transformations \cite{NHQC2,NHQC8,NHQC10}, since the evolution of logical-qubit subspace satisfies the parallel-transport condition, i.e.,
\begin{eqnarray}
\label{Eq9}
L_1\mathcal{H}_1L_1=U_1(t)L_1\mathcal{H}_1L_1U^{\dagger}_1(t)=0
\end{eqnarray}
where $L_1=|0\rangle_{L}\langle 0|+|1\rangle_{L}\langle 1|$ is the projection operator; and cycle condition, i.e.,
\begin{eqnarray}
\label{Eq10}
S_1(\tau)&=&\textrm{Span}\{U_1(\tau) |0\rangle_L, U_{1}(\tau)|1\rangle_L \}\notag\\
&=&\textrm{Span}\{|0\rangle_L, |1\rangle_L \}=S_1.
\end{eqnarray}

However, in the practical physical implementation, the performance of the proposed single-logical-qubit gate $U_{L_1}(\tau)$ is inevitably limited by the decoherence effect of the target quantum system. Therefore, we here consider the effects of decoherence and the high-order oscillating terms in the logical-qubit subspace by numerically simulating the Lindblad master equation of
\begin{eqnarray}
\label{Eq11}
\dot\rho_1&=& -\text{i}\left[\mathcal{H}_S, \rho_1\right] \notag\\
&+&\sum_{l=\textrm{a},1,2}\left\{ \sum_{k=1}^2 \left(\frac {\kappa^l_-} {2}\mathscr{L}( \lambda_k\sigma^k_l)+ \frac {\kappa^l_z} {2}\mathscr{L}( k\chi^k_l)\right)\right\},\quad
\end{eqnarray}
where $\rho_1$ is the reduced density matrix of the considered quantum system, $\mathscr{L}(\mathcal{A})=2\mathcal{A}\rho_1
\mathcal{A}^\dagger-\mathcal{A}^\dagger \mathcal{A} \rho_1 -\rho_1 \mathcal{A}^\dagger \mathcal{A}$ is the Lindbladian of the operator $\mathcal{A}$, and $\kappa^l_-$ and $\kappa^l_z$ are the relaxation and dephasing rates of the \textit{l}th transmon, respectively. We next choose the NOT and Hadamard holonomic gates as two typical examples to fully evaluate their gate performances.
For a general initial state $|\psi_1\rangle=|0\rangle_{\textrm{a}}(\cos\theta_1|0\rangle_L+\sin\theta_1|1\rangle_L)$, the ideal final state is $|\psi_{f_{k=\textrm{N},\textrm{H}}}\rangle= U_{1}(\tau_k)|\psi_1\rangle$, we use gate fidelity $F_{k}^\textrm{G}=\frac {1} {2\pi}\int_0^{2\pi} \langle \psi_{f_k}|\rho_1|\psi_{f_k}\rangle d\theta_1$ \cite{gatefidelity} to quantify the gate performance, where the integration is numerically done for 1001 input states with $\theta_1$ being uniformly distributed over $[0, 2\pi]$.
According to the current state-of-art of experiments \cite{Martinis1,Martinis2}, the parameters are set to be $\kappa=\kappa^1_-=\kappa^1_z=\kappa^2_-=\kappa^2_z=\kappa^\textrm{a}_-=\kappa^\textrm{a}_z=2\pi\times 4$ kHz, $\alpha_1=2\pi\times 220$ MHz, $\alpha_2=2\pi\times 180$ MHz, $\alpha_\textrm{a}=2\pi\times 210$ MHz and $g_{1\textrm{a}}=g_{2\textrm{a}}=2\pi\times 12$ MHz.
Meanwhile, for the NOT and Hadamard gates, we modulate the qubit-driving parameters $J_1(\beta_2)/J_1(\beta_1)= 1$ and $0.414$ to ensure $\theta=\pi/2$ and $\pi/4$, respectively. In Figs. \ref{Fig2}(a) and \ref{Fig2}(b), our numerical simulation shows that, when the qubit frequency differences $\Delta_1, \Delta_2$ are tuned, through tune the qubit frequency of the auxiliary, within the range of $ 2\pi\times(335\pm2)$ MHz, the gate fidelities of the NOT and Hadamard holonomic gates can both approach to $99.80\%$.
We can also evaluate these gates by state populations and the state fidelities defined by $F_{k}=\langle\psi_{f_{k}}|\rho_1|\psi_{f_{k}}\rangle$. Here, we assume that the initial state of the quantum system is $|\psi_1\rangle=|0\rangle_{\textrm{a}}|0\rangle_L$, then the NOT and Hadamard gates should be the result in the ideal final states $|\psi_{f_{\textrm{N}}}\rangle=|0\rangle_{\textrm{a}}|1\rangle_L$ and $|\psi_{f_{\textrm{H}}}\rangle=|0\rangle_{\textrm{a}}(|0\rangle_L+|1\rangle_L)/\sqrt{2}$, respectively.
As shown in Figs. \ref{Fig2}(c) and \ref{Fig2}(d), the state fidelities of the NOT and Hadamard gates can be also as high as $F_{\textrm{N}}=99.86\%$ and $F_{\textrm{H}}=99.75\%$, respectively, where we note that the initial and final states of the auxiliary qubit will both be in its ground state.

As the single-logical-qubit gate manipulations are limited in the single-excitation subspace of the original Hamiltonian $\mathcal{H}_S$, the leakage to the multi-excitation subspaces can be eliminated, and thus the gate infidelity will originate from the effects of decoherence and the high-order oscillating terms, i.e.,  $\sum_{n\neq1} \sum_{l=1,2} J_{n}(\beta_l)g_{l\textrm{a}}e^{\text{i}(1-n)\Delta_l} |10\rangle_{l\textrm{a}}\langle 01|+ \mathrm{H.c.}$. Note that our numerical simulation is based on $\mathcal{H}_S$ without any approximation, thus it can verify our analytical results and be used to quantify the contribution of different error sources to the gate infidelity. Through our analysis, we find that the decoherence produces about $0.1\%$ gate infidelity, and the remaining infidelity is due to the high-order oscillating terms. In addition, for the case of NOT gate, since the effective coupling strengths of the two pairs of parametrically coupled transmons are equal, i.e., $J_1(\beta_2)/J_1(\beta_1)= 1$, so the effects of high-order oscillating terms on them are the same, and thus the gate fidelities in Fig. \ref{Fig2}(a) are symmetric with respect to $\Delta_1$ and $\Delta_2$.
However, for the Hadamard gate, as $J_1(\beta_2)/J_1(\beta_1)= 0.414$, the effective coupling strength between transmons $\textrm{T}_1$ and $\textrm{T}_a$ is stronger, thus the same oscillating frequency term will introduce more error on the $\textrm{T}_1$ and $\textrm{T}_a$ pair, which explains why the gate infidelities are asymmetrical with respect to $\Delta_1$ and $\Delta_2$, as shown in Fig. \ref{Fig2}(b).

\section{Two-logical-qubit holonomic gates}
We next consider the implementation of the two-logical-qubit controlled-NOT holonomic gate, which is a nontrivial element in constructing universal quantum gates. As shown in Fig. \ref{Fig1}(a), we treat two pairs of transmon qubits, e.g., $\textrm{T}_1$ and $\textrm{T}_2$, $\textrm{T}_3$ and $\textrm{T}_4$, coupled to the same auxiliary transmon $\textrm{T}_\textrm{a}$, as two logical units on the 2D square lattice to encode the first and second DFS logical qubits, respectively. There exists a four-dimensional DFS
\begin{eqnarray}
\label{Eq12}
S_2= \textrm{Span}\{&&|1010\rangle_{1234}=|00\rangle_L,|1001\rangle_{1234}=|01\rangle_L \notag\\
                    &&|0110\rangle_{1234}=|10\rangle_L,|0101\rangle_{1234}=|11\rangle_L\}.\ \ \ \
\end{eqnarray}

For the two-logical qubit case, we find that the general evolution of two-logical-qubit states can be completed by making transmon $\textrm{T}_2$ synchronously achieve the parametrically tunable coupling interaction with transmons $\textrm{T}_3$ and $\textrm{T}_4$, by introducing the qubit-frequency drivings. The corresponding control Hamiltonian is $\mathcal{H}_T=\mathcal{H}_{23}+\mathcal{H}_{24}$. Different from the processing of a single-logical qubit, we here modulate the qubit-frequency driving parameters $\nu_3=\Delta_3-\alpha_2$, $\phi_3=\varphi+\pi/2$ and $\nu_4=\Delta_4-\alpha_2$, $\phi_4=-\pi/2$ with $\Delta_l=\omega_2-\omega_l$. The obtained resonant interaction Hamiltonian reads as
\begin{eqnarray}
\label{Eq13}
\mathcal{H}_2=g'_{23}e^{\text{i}\varphi}|11\rangle_{23}\langle 20| -g'_{24}|20\rangle_{24}\langle 11|+ \mathrm{H.c.},
\end{eqnarray}
where $g'_{2l}=\sqrt{2}J_1(\beta_l) g_{2l}$. By setting $\Omega=\sqrt{g'^2_{23}+g'^2_{24}}$ and $\vartheta=2\tan^{-1}(g'_{23}/ g'_{24})$, the reduced Hamiltonian is
\begin{eqnarray}
\label{Eq14}
\mathcal{H}_{L_2}=\Omega\left(
\begin{array}{ccc}
 0 & 0 & \sin\frac{\vartheta} {2}e^{\text{i}\varphi} \\
 0 & 0 & -\cos\frac{\vartheta} {2} \\
 \sin\frac{\vartheta} {2}e^{-\text{i}\varphi} & -\cos\frac{\vartheta} {2} & 0
\end{array}
\right)
\end{eqnarray}
in the two-logical-qubit subspace $\{|10\rangle_L, |11\rangle_L, |a\rangle_L\}$, where $|a\rangle_L=|0200\rangle_{1234}$ is used as an ancillary state.

\begin{figure}[tbp]
  \centering
  \includegraphics[width=0.95\linewidth]{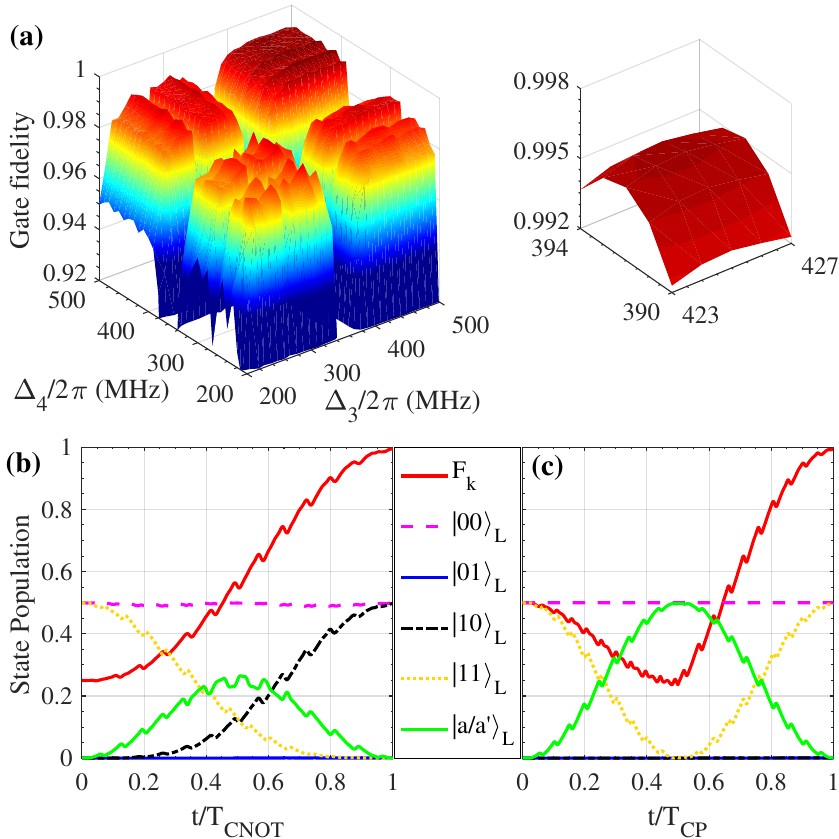}
  \caption{(a) The gate fidelities as functions of the qubit frequency differences $\Delta_3$ and $\Delta_4$, for the controlled-NOT holonomic gate. State populations and the state fidelity dynamics of the controlled-NOT (b) and controlled-phase (c) gates with the initial state being $\frac{1} {\sqrt{2}}(|00\rangle_L+|11\rangle_L)$.}\label{Fig3}
\end{figure}

When we choose $\Omega T=\pi$ at the final time $T$, within the two-logical-qubit subspace $\{|00\rangle_L, |01\rangle_L, |10\rangle_L, |11\rangle_L\}$, the evolution operator can be expressed as
\begin{eqnarray}
\label{Eq15}
U_{L_2}(T)=\left(
\begin{array}{cccc}
 1 & 0 & 0 & 0 \\
 0 & 1 & 0 & 0 \\
 0 & 0 & \cos\vartheta & \sin\vartheta e^{\text{i}\varphi} \\
 0 & 0 & \sin\vartheta e^{-\text{i}\varphi} & -\cos\vartheta
\end{array}
\right).
\end{eqnarray}

Similar to the single-logical-qubit case, the above obtained evolution operator is of a holonomic nature, since the following parallel-transport and cyclic evolution conditions can be satisfied, i.e.,
\begin{eqnarray}
\label{EqC1}
L_2\mathcal{H}_{2}L_2=U_{L_2}(t)L_2\mathcal{H}_{2}L_2U^{\dagger}_{L_2}(t)=0,
\end{eqnarray}
and
\begin{eqnarray}
\label{EqC2}
S_2(T) &=& \textrm{Span}\{U_{L_2}(T)|00\rangle_L,U_{L_2}(T)|01\rangle_L,        \notag \\
&&\quad \quad U_{L_2}(T)|10\rangle_L,U_{L_2}(T)|11\rangle_L\} \notag \\
&=&\textrm{Span}\{|00\rangle_L,|01\rangle_L,|10\rangle_L,|11\rangle_L\}\notag \\
&=&S_2,
\end{eqnarray}
where $L_2=|00\rangle_{L}\langle 00|+|01\rangle_{L}\langle 01|+|10\rangle_{L}\langle 10|+|11\rangle_{L}\langle 11|$ is the projection operator of two-logical qubit.

In this way, the nontrivial two-logical-qubit holonomic gate can be obtained. For example, by modulating qubit-driving parameters $\varphi=0$ and $J_1(\beta_3)/J_1(\beta_4)=1$ to ensure $\vartheta=\pi/2$, the two-logical-qubit controlled-NOT holonomic gate can be achieved, which is a nontrivial entangling gate, together with an arbitrary single-qubit gate,  they constitute a universal set of quantum gates.

Next, to fully evaluate the performance of the two-logical-qubit gate, for the general initial state $|\psi_2\rangle=(\cos\vartheta_1|0\rangle_L+\sin\vartheta_1|1\rangle_L)\otimes
(\cos\vartheta_2|0\rangle_L+\sin\vartheta_2|1\rangle_L)$, we here define the gate fidelity of two-logical qubit as
\begin{eqnarray}
\label{Eq16}
F^\textrm{G}_{\textrm{CNOT}}=\frac {1} {4\pi^2}\int_0^{2\pi} \int_0^{2\pi} \langle \psi_{f_{\textrm{CNOT}}}|\rho_2|\psi_{f_{\textrm{CNOT}}}\rangle d\vartheta_1d\vartheta_2
\end{eqnarray}
with $|\psi_{f_{\textrm{CNOT}}}\rangle=U_{L_2}(T_{\textrm{CNOT}})|\psi_2\rangle$ being the ideal final state. Set parameters $\alpha_3=2\pi\times 220$ MHz, $\alpha_4=2\pi\times 200$ MHz, $g_{23}=g_{24}=2\pi\times 7$ MHz, and a uniform decoherence rate $\kappa=\kappa^1_-=\kappa^1_z=\kappa^2_-=\kappa^2_z=\kappa^3_-=\kappa^3_z=\kappa^4_-=\kappa^4_z=2\pi\times 4$ kHz. By analyzing the numerical simulation results as shown in Fig. \ref{Fig3}(a), when the qubit frequency differences $\Delta_{3}\in2\pi\times(392\pm2)$ and $\Delta_{4}\in2\pi\times(425\pm2)$, and qubit-driving frequencies are modulated to $\nu_3=\Delta_3-\alpha_2$ and $\nu_4=\Delta_4-\alpha_2$, the gate fidelities of the controlled-NOT holonomic gate is $99.55\%$,
where the decoherence effect produces about $0.2\%$ gate infidelity, and the remaining gate infidelity originates from  the presence of high-order oscillating terms in the two-logical-qubit subspace and the leakage to the non-two-logical-qubit subspaces. The corresponding state fidelity is $99.52\%$ when the initial state is $\frac{1} {\sqrt{2}}(|00\rangle_L+|11\rangle_L)$, as shown in Fig. \ref{Fig3}(b). In addition, we can find that, if the driving frequency does not satisfy the corresponding constraints to achieve the form of effective resonant Hamiltonian $\mathcal{H}_2$ in Eq. (\ref{Eq13}), the effects of both high-order oscillating terms  and the leakage to the non-two-logical-qubit subspaces will become severe, which explains why the gate fidelities are not good in some parameter regions in Fig. \ref{Fig3}(a).

\begin{figure}[tbp]
  \centering
  \includegraphics[width=0.8\linewidth]{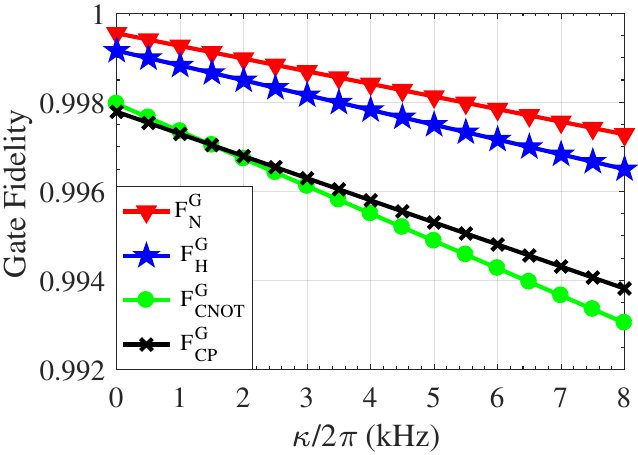}
  \caption{Gate fidelities with respect to the different uniform decoherence rate $\kappa$.}\label{Fig4}
\end{figure}

\section{Discussion and Conclusion}

The above two-logical qubit constitutes a cross-type unit to implement universal NHQC with the DFS
encoding on a 2D square superconducting transmon-qubit lattice. In addition, based on the scalability of our scheme as shown in Fig. \ref{Fig1}(a), the two-logical qubit can be arranged not only in the horizontal direction but also in the vertical and diagonal directions. Such as the case of the two-logical qubit arranged in the horizontal direction, we can also use the coupling interaction between the two neighboring transmons, i.e., $\textrm{T}_2$ and $\textrm{T}_{4'}$, to induce our wanted two-logical-qubit controlled-phase holonomic gate, where a pair of transmons $\textrm{T}_{3'}$ and $\textrm{T}_{4'}$ are a new unit to encode the second DFS logical qubit.

The control Hamiltonian can be defined as $\mathcal{H}'_T=\mathcal{H}_{24'}$, where transmon $\textrm{T}_2$ is introduced additional qubit-frequency driving in the form of $\omega'_{2}(t)=\omega_2+\varepsilon'_{2} \sin(\nu'_{2} t+\varphi_{2})$. We here modulate the parameter $\nu'_2=\omega_2-\omega_{4'}-\alpha_2$,
the effective resonant interaction Hamiltonian can be obtained as
\begin{eqnarray}
\label{EqCPH}
\mathcal{H}'_{L_2}=g'_{24'}e^{\text{i}(\varphi_{2}+\frac{\pi} {2})}|a'\rangle_{L}\langle 11|+ \mathrm{H.c.},
\end{eqnarray}
where $|a'\rangle_L=|0200\rangle_{123'4'}$ is used as an ancillary state, and $g'_{24'}=\sqrt{2}J_1(\beta'_2) g_{24'}$ with $\beta'_2=\varepsilon'_{2}/\nu'_2$.
Then, by setting $g'_{24'}T_{\textrm{CP}}=\pi$, at the final time $T_{\textrm{CP}}$, within the two-logical-qubit subspace $\{|00\rangle_L, |01\rangle_L, |10\rangle_L, |11\rangle_L\}$, the controlled-phase gate $U_{\mathrm{CP}}=\mathrm{diag}\{1,1,1, e^{i\xi} \}$ can be obtained, where the non-Abelian geometric phase $\xi$ is achieved by changing the phase $\varphi_{2}$ to $\varphi_{2}+\pi+\xi$ at the middle moment $T_{\textrm{CP}}/2$ in a single-loop scenario.
The proof of the holonomic nature of the controlled-phase gate is similar to that of in Eqs. (\ref{EqC1}) and Eq. (\ref{EqC2}).
Taking the case of $\xi=\pi/2$ as an example, the gate fidelity of the controlled-phase gate can reach $99.60\%$ by setting the qubit frequency difference of the two transmons $\textrm{T}_2$ and $\textrm{T}_{4'}$ in the range of $(420\pm2)$ MHz, under the parameters of transmon $g_{24'}=2\pi\times 7$ MHz, $\alpha_{4'}=2\pi\times 200$ MHz,
and a uniform decoherence rate $\kappa=\kappa^1_-=\kappa^1_z=\kappa^2_-=\kappa^2_z=\kappa^3_-=\kappa^3_z=\kappa^4_-=\kappa^4_z=2\pi\times 4$ kHz.
The state dynamics details are shown in Fig. \ref{Fig3}(c), the state fidelity is $99.46\%$ when the initial state is $\frac{1} {\sqrt{2}}(|00\rangle_L+|11\rangle_L)$. In this case, we can find that the decoherence produces about $0.2\%$ gate infidelity, and the remaining  infidelity is due to the high-order oscillating terms and leakage to the non-logical-qubit subspaces.

Finally, to intuitively measure the effects of decoherence on the holonomic quantum gates, we depict the trend of gate fidelities under the uniform decoherence rate $\kappa/2\pi\in [0,8]$ kHz in Fig. \ref{Fig4}, which further verifies the feasibility of the physical realization of our scheme.

In conclusion, we have proposed to implement scalable universal NHQC with DFS encoding on a 2D square superconducting transmon-qubit lattice, in a tunable and all-resonant way, avoiding complexity of experimental manipulation and level leakage to multi-excitation subspaces. Thus, our scheme provides a promising way towards the practical realization of high-fidelity NHQC.

\acknowledgments

This work was supported by Key-Area Research and Development Program of GuangDong Province (Grant No. 2018B030326001), the National Natural Science Foundation of China (Grant No. 11874156), and the National Key R\&D Program of China (Grant No. 2016YFA0301803).

L.-N. Ji and T. Chen contributed equally to this work.

\end{document}